\documentclass[10pt]{article}
\usepackage{multirow}
\usepackage[dvips]{graphicx}

\usepackage[psamsfonts]{amssymb}
\usepackage{amsxtra}
\usepackage{threeparttable}
\usepackage{amsmath}
\usepackage{amssymb}

\usepackage{graphicx}

\usepackage{cite}

\usepackage{color}

\usepackage{setspace}
\usepackage{float}

\usepackage{url}

\usepackage[labelfont=bf,labelsep=period,justification=raggedright]{caption}

\makeatletter
\renewcommand{\@biblabel}[1]{\quad#1.}
\makeatother

\date{May 15, 2013}

\begin{document}

\begin{flushleft}
{\Large
\textbf{EEG Signal Processing and Classification for the Novel Tactile--Force Brain--Computer Interface Paradigm}\footnote{The final publication is available at IEEE Xplore \url{http://ieeexplore.ieee.org} and the copyright of the final version has been transferred to IEEE \copyright2013}

}
Shota Kono$^1$, Daiki Aminaka$^1$, Shoji Makino$^1$,
and Tomasz M. Rutkowski$^{1,2,}$\footnote{The corresponding author. E-mail: \url{tomek@tara.tsukuba.ac.jp}}
\\
\bf{$^1$}Life Science Center of TARA, University of Tsukuba, Tsukuba, Japan\\
\bf{$^2$}RIKEN Brain Science Institute, Wako-shi, Japan\\
E-mail: \url{tomek@tara.tsukuba.ac.jp}\\
\url{http://bci-lab.info/}
\end{flushleft}

\section*{Abstract}

The presented study explores the extent to which tactile--force stimulus delivered to a hand holding a joystick can serve as a platform for a brain computer interface (BCI). The four pressure directions are used to evoke tactile brain potential responses, thus defining a tactile--force brain computer interface (tfBCI). We present brain signal processing and classification procedures leading to successful interfacing results. Experimental results with seven subjects performing online BCI experiments provide a validation of the hand location tfBCI paradigm, while the feasibility of the concept is illuminated through remarkable information-transfer rates.

\noindent{\bf Keywords:} EEG; tactile BCI; brain signal processing; brain somatosensory evoked response.

\section{Introduction}

The state--of--the--art BCI is usually based on mental visual and motor imagery paradigms, which require substantial user training and good eyesight from the subject~\cite{bciBOOKwolpaw}. Alternative solutions have been proposed recently to make use of spatial auditory~\cite{iwpash2009tomek} or tactile (somatosensory) modalities~\cite{sssrBCI2006,tactileBCIwaiste2010,JNEtactileBCI2012,HiromuBCImeeting2013,tactileAUDIOvisualBCIcompare2013} to enhance brain-computer interfacing comfort. The concept proposed and reported based on a conducted pilot study in this paper further extends the previously reported by the authors~\cite{HiromuBCImeeting2013,tomekHAID2013} brain somatosensory (tactile) channel to allow for tactile--force based stimulus application. The rationale behind the use of the tactile--force tactile channel is that it is usually far less loaded and more intuitive to learn comparing with auditory or even visual modality interfacing applications. A very recent report~\cite{tactileAUDIOvisualBCIcompare2013} additionally has confirmed superiority of the tactile BCI in comparison with visual and auditory modalities tested with a locked--in syndrome (LIS) subject~\cite{alsTLSdiagnosis1966}. 

A recent report~\cite{JNEtactileBCI2012} proposed to utilize as the tactile BCI a Braille--code stimulator with $100$~ms static force push stimulus delivered to each of six fingers to evoke a somatosensory evoked potential (SEP) response and the following P300 attentional modulation related peak in an event--related potential (ERP).
The P300 response is a positive electroencephalogram ERP deflection starting at around $300$~ms and lasting for $200-300$~ms after an expected stimulus in an oddball (random) series of distractors~\cite{bciBOOKwolpaw}. 
Examples of averaged P300 responses are depicted using red lines with standard errors in Figure~\ref{fig:EEGerp} and in form of color coded diagrams in Figure~\ref{fig:alleegauc}.

The P300 brain response is considered to be the most reliable and easy to capture from EEG in majority of human subjects. Thus, the P300 is commonly used in BCI applications~\cite{bciBOOKwolpaw,bci2000book}.

This paper reports on the novel successful application of the tactile--force BCI. We present very encouraging results obtained with seven healthy subjects of whom the majority scored with $100\%$ accuracy in online BCI experiments.

The rest of the paper from now on is organized as follows. The next section introduces the materials and methods used in the tactile--force BCI study. It also outlines the experiments conducted. The results obtained in EEG online experiments with seven healthy BCI subjects are then discussed. Finally, conclusions are formulated and directions for future research are outlined.

\section{Materials and Methods}

The experiments in the reported study involved seven healthy subjects (six males and one female; mean age of $24.71$ years, with a standard deviation of $7.5$ years). All the experiments were performed at the Life Science Center of TARA, University of Tsukuba, Japan. The online EEG BCI experiments were conducted in accordance with \emph{The World Medical Association Declaration of Helsinki - Ethical Principles for Medical Research Involving Human Subjects}. The psychophysical and EEG recording for BCI paradigm experimental procedures were approved by the Ethical Committee of the Faculty of Engineering, Information and Systems at University of Tsukuba, Tsukuba, Japan. The participants agreed voluntarily to take part in the experiments. The details of the tactile--force stimulus creation, psychophysical and EEG experimental protocols are described in the following subsections. 

\subsection{Tactile--Force Stimulus}

\begin{figure}
	\centering
	\includegraphics[width=\linewidth]{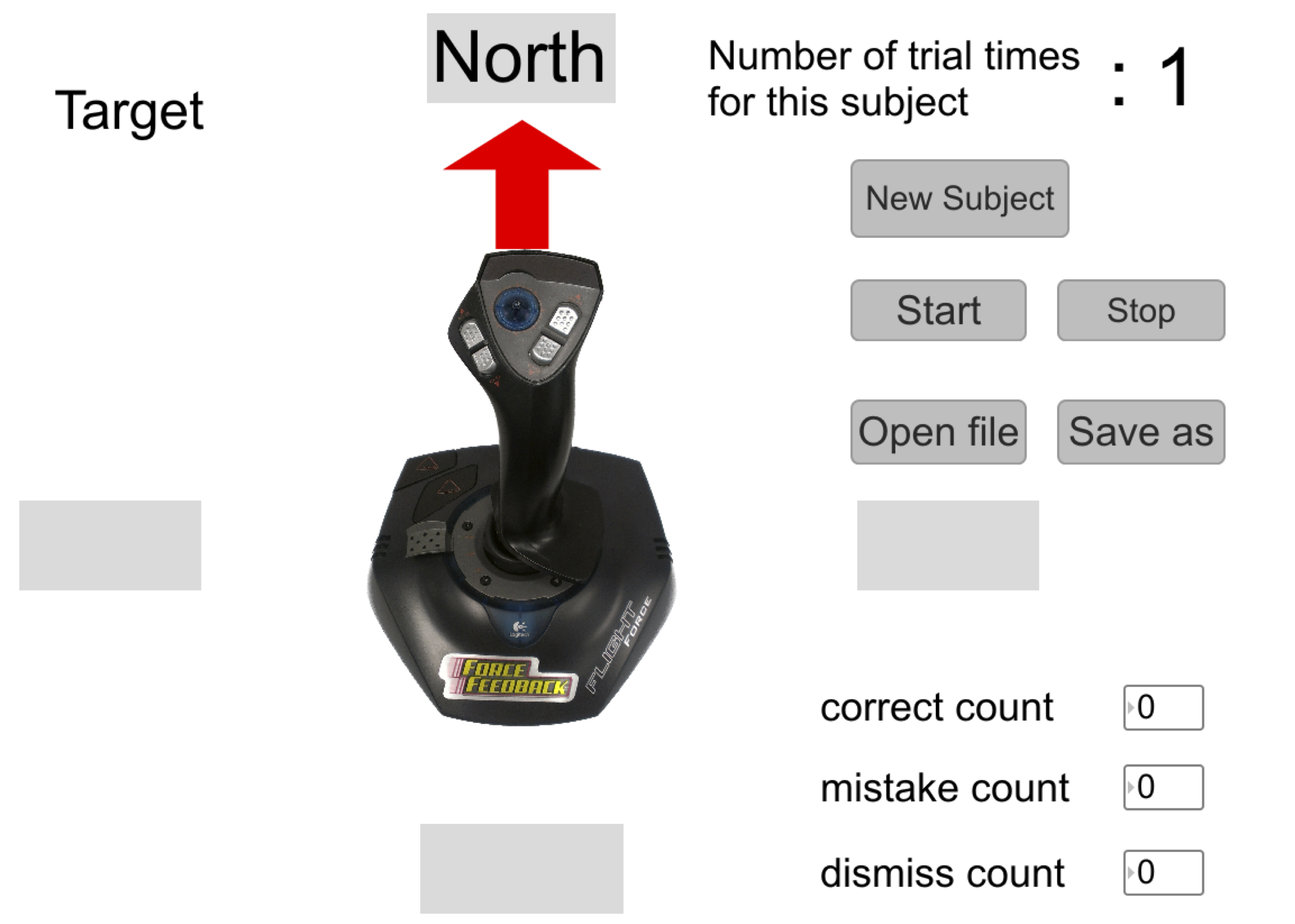}
	\caption{The visual instruction screen presented to the subjects during the psychophysical experiment developed in \textsf{MAX~6}~\cite{maxMSP}.}
	\label{fig:MAX}
\end{figure}

The tactile stimuli were delivered as movements generated by a portable computer in \textsf{MAX~6}~\cite{maxMSP} environment as depicted in form of visual interface with instructions to the subject in Figure~\ref{fig:MAX}.
Each tactile stimulus was generated via a Force Feedback Joystick Driver for Java~\cite{javajoystickdriver}.
The stimuli were delivered to the subject's right palm via the \textsf{FLIGHT FORCE} joystick by Logitech.

\begin{figure}
	\centering
	\includegraphics[width=0.6\linewidth]{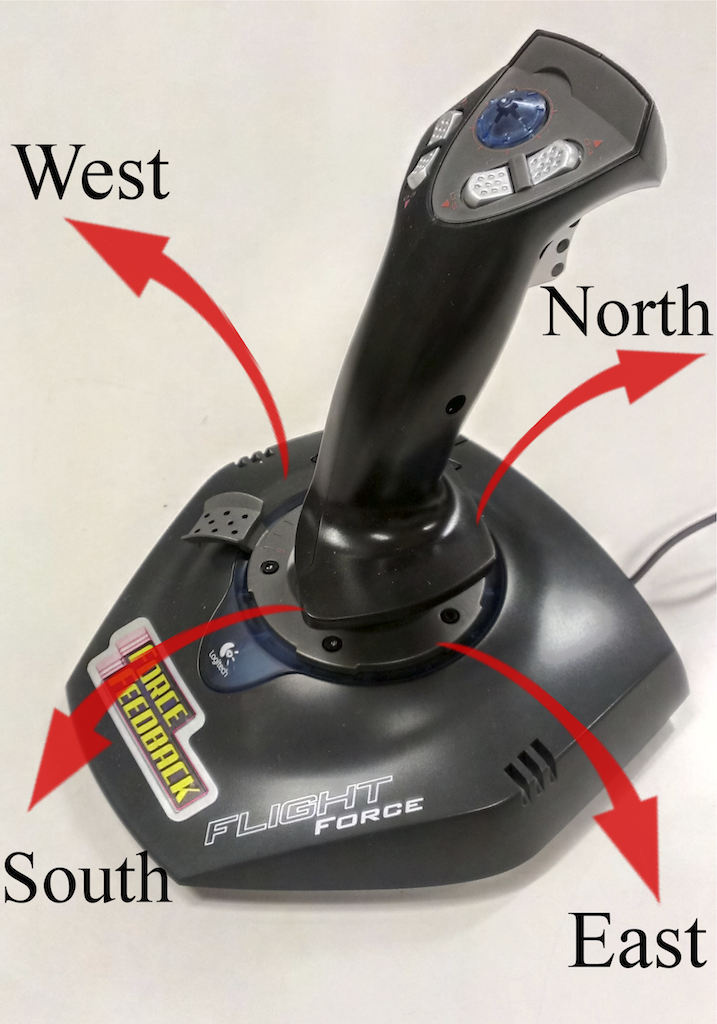}
	\caption{The force--feedback (or tactile--force) joystick \textsf{FLIGHT FORCE} by Ligitech used in experiments reported in this paper. The tactile--force stimulus was delivered to the subject's dominant hand. Four movements, defined as \emph{North, East, South} and \emph{West} directions, were executed randomly from a computer causing the joystick handle to move and push the subject's hand automatically.}\label{fig:JoystickMovement}
\end{figure}

There were four stimulus tactile--force direction patterns delivered in random order to the subject hands. The directions were labeled as \emph{North, East, West,} and \emph{South} as depicted in Figure~\ref{fig:JoystickMovement}. For example, the \emph{North} directions stimulus interaction caused the joystick generate a forward tactile--force pressure on the subject's hand holding it. Similarly the \emph{South, East,} and \emph{West} stimulus directions were causing backward, right, and left tactile--force pressures on the subject hand respectively. The joystick returned to the center position (no pressure) after the each presented stimulus after the presentation time of $100$~ms (see Tables~\ref{tab:Psychoconditions}~and~\ref{tab:EEGconditions} with experimental condition details summarized).

During the both psychophysical and EEG experiments the subject held the joystick handle using a dominant hand (right in case of all the subjects participating in this study) and responded (button press in psychophysical-- and mental confirmation/counting in case of EEG--experiment) only to the instructed direction. The instruction which directions to attend were presented visually using the same \textsf{MAX~6} environment program that created the stimulus and communicated it via the Java driver to joystick as depicted in Figure~\ref{fig:MAX}.

\subsection{Tactile--Force Psychophysical Experiment Protocol}

The psychophysical experiment was conducted to investigate the stimulus tactile--force direction influence on the subject behavioral response time and accuracy.
The behavioral responses were collected using a trigger button on the joystick handle and the \textsf{MAX~6} program. The subject was instructed which stimulus to attend in each session by an arrow on the computer display pointing the direction of \emph{an target} as depicted in Figure~\ref{fig:MAX}.
In the each psychophysical experiment the subject was presented with $80$ \emph{target} and $240$ \emph{non--target} directions as stimuli.

Each trial was composed of $100$~ms tactile--force pressures delivered to subject hand in randomized order with an inter--stimulus--interval (ISI) of $900$~ms. Every random sequence thus contained a single \emph{target} and three \emph{non--targets}. A single session was composed of the ten trials for each tactile--force \emph{target}. The choice of the relatively long ISI was justified by a slow behavioral response in comparison to the EEG evoked potential, as described in the next section. The tactile--force psychophysical experiment protocol details are summarized in Table~\ref{tab:Psychoconditions}.

\begin{table}
	\caption{Tactile--force psychophysical experiment protocol conditions and details}\label{tab:Psychoconditions}
	\begin{center}
	\begin{tabular}{|l|l|}
	\hline
	Condition							& Detail \\
	\hline \hline
	Number of subjects					& $7$ \\
	Tactile stimulus length				& $100$~ms \\
	Inter--stimulus--interval (ISI)		& $900$~ms \\
	Stimulus generation					& \textsf{FLIGHT FORCE} joystick \\
										& by Logitech \\
	Number of trials for each subject 	& $10$ \\
	\hline
	\end{tabular}
	\end{center}
\end{table}

The behavioral response times were registered with the same \textsf{MAX~6} program, also used for the stimulus generation and instruction presentation as depicted in Figure~\ref{fig:MAX}. The goal of the psychophysical experiment, investigating behavioral response times and \emph{target} recognition accuracy in order to test even distribution of cognitive loads (tasks difficulties) among the four tactile--force stimuli, was reached and the results are discussed in the Section~\ref{sec:psychoRESULTS}.

\subsection{EEG tfBCI Experiment Protocol}

In the BCI experiments EEG signals were captured with a portable EEG amplifier system \textsf{g.USBamp} by g.tec Medical Instruments, Austria. Sixteen active wet EEG electrodes were used to capture brain waves with event related potentials (ERP) with attentional modulation elucidated by the so--called ``aha--'' or P300--response~\cite{bciBOOKwolpaw}. 
The EEG electrodes were attached to the head locations \emph{Cz, CPz, P3, P4, C3, C4, CP5, CP6, P1, P2, POz, C1, C2, FC1, FC2,} and \emph{FCz,} as in $10/10$ intentional system~\cite{Jurcak20071600}.
A reference electrode was attached to a left mastoid and a ground electrode on the forehead at \emph{FPz} position respectively.  
No electromagnetic or electromyographic (EMG) interference was observed from the moving joystick.
Details of the EEG experimental protocol are summarized in Table~\ref{tab:EEGconditions}.

\begin{table}
	\caption{Conditions and details of the tfBCI EEG experiment}\label{tab:EEGconditions}	
	\begin{center}
	\begin{tabular}{|l|l|}
	\hline
	Condition								& Detail \\
	\hline \hline
	Number of subjects						& $7$ \\ 
	Tactile stimulus length					& $100$~ms \\ 
	Inter--stimulus--interval (ISI)			& $900$~ms \\ 
	EEG recording system					& gUSBamp active wet EEG \\
											& electrodes system \\
	Number of the EEG channels				& $16$ \\
	EEG electrode positions					& Cz, CPz, P3, P4, C3, C4, CP5, \\
											& CP6, P1, P2, POz, C1, C2, FC1, \\
											& FC2, FCz \\ 
	Reference electrode						& Behind the subject's left ear \\
	Ground electrode						& On the forehead (FPz) \\
	Stimulus generation						& \textsf{FLIGHT FORCE} joystick \\
											& by Logitech \\
	Number of trials for each subject		& $10$ \\
	\hline
	\end{tabular}
	\end{center}
\end{table}

The EEG signals were recorded and preprocessed by a \textsf{BCI2000}--based application~\cite{bci2000book}, using a stepwise linear discriminant analysis (SWLDA) classifier~\cite{krusienski2006} with features drawn from the $0-800$~ms ERP interval.
The sampling rate was set to $256$~Hz, the high pass filter at $0.1$~Hz, and the low pass filter at $40$~Hz. The ISI was $300$~ms and each tactile--force stimulus duration was $100$~ms. 

Instructions to the subject which tactile--force stimulus direction to attend were presented visually as in the previous psychophysical experiments using the \textsf{MAX~6} program as depicted in Figure~\ref{fig:MAX}. Each \emph{target} was presented ten times in a random series with the remaining \emph{non--targets}. A procedure  of ten single ERP responses averaging was used in order to enhance the P300 response in a very noisy EEG~\cite{bci2000book,krusienski2006}.

\section{Results}

This section presents and discusses results that we obtained in the psychophysical and in the online tfBCI experiments. The very encouraging results obtained in the psychophysical and tfBCI paradigm experiment support the proposed concept of tactile--force modality.

\subsection{Tactile--force Psychophysical Experiment Results}\label{sec:psychoRESULTS}

\begin{figure}
	\centering
	\includegraphics[width=\linewidth]{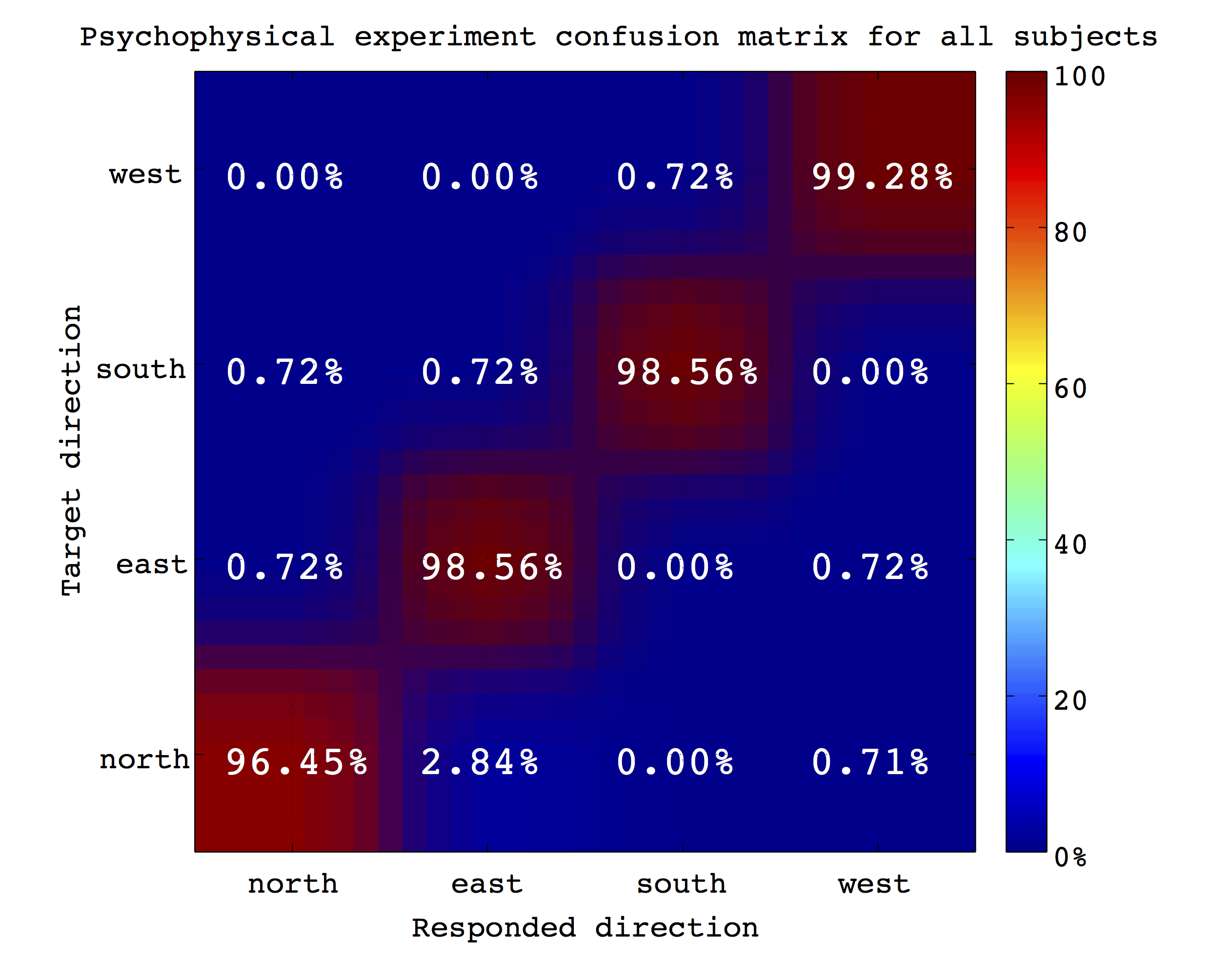}
	\caption{Tactile--force interface psychophysical experiment results in form of a confusion matrix of the grand mean averaged subject accuracy results. Rows of the above matrix denote the instructed \emph{targets} and columns the subject response. A diagonal of the matrix visualizes the correct response, while the off--diagonal values the subject errors. Numerical percentage values represent the response accuracies. In the contacted experiments the mean errors were marginal (below one percent). There were also no systematic errors observed (common mistakes between pairs of patterns), which further validated the tactile--force stimulus design.}\label{fig:confMX}
\end{figure}

\begin{figure}
	\centering
	\includegraphics[width=0.8\linewidth]{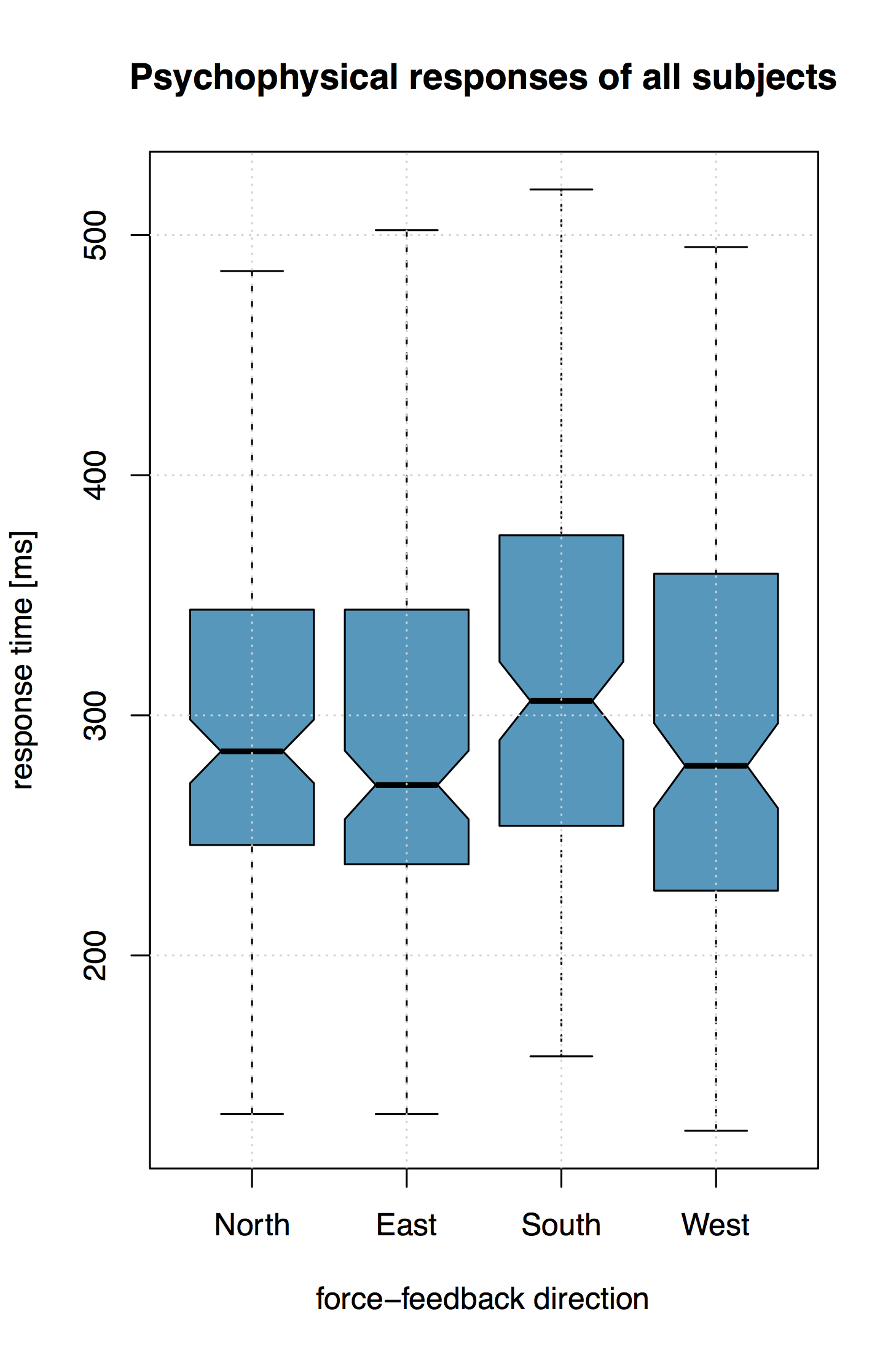}
	\caption{Boxplots of the tactile--force psychophysical experiment response time distributions of the four \emph{North, East, South} and \emph{West} joystick directions. The differences among median were not significantly different (as tested with pairwise \emph{Wilcoxon} statistical test). The boxplots depict also response time interquartile ranges (edges of the boxes) of the response time distributions, which almost completely cover each other in the above plot.}\label{fig:responsetime}
\end{figure}

The psychophysical experiment accuracy results are summarized in Table~\ref{tab:psychoaccuracy}, depicted in form of a confusion matrix in Figure~\ref{fig:confMX}, and as boxplot response time distributions in Figure~\ref{fig:responsetime}, where the median response times and the interquartile ranges are depicted for each direction respectively (see also Figure~\ref{fig:MAX} for the directions).

This result confirmed the stimulus similarity since the behavioral responses for all the directions were basically the same. This finding validated the design of the following tfBCI EEG experiment, since the four tactile--force patterns resulted with similar cognitive loads as confirmed by the same accuracies and response times.

\begin{table}
	\caption{Psychophysical experiment results (note, this is not a binary accuracy case yet the one with a theoretical chance level of $25\%$) in tactile--force interface task.}\label{tab:psychoaccuracy}
	\begin{center}
	\begin{tabular}{|c|c|}
	\hline 
	Subject number 	& The best psychophysical accuracy \\
	\hline \hline
	$\#1$			& $100\%$ \\
	$\#2$			& $100\%$ \\
	$\#3$			& $100\%$ \\
	$\#4$			& $~95\%$ \\
	$\#5$			& $100\%$ \\
	$\#6$			& $100\%$ \\
	$\#7$			& $100\%$ \\
	\hline
	{\bf Average:}	& $\bf 99.3\%$ \\
	\hline
	\end{tabular}
	\end{center}
\end{table}

\subsection{Online EEG Tactile--Force BCI Experiment Results}

\begin{figure}
	\centering
	\includegraphics[width=\linewidth]{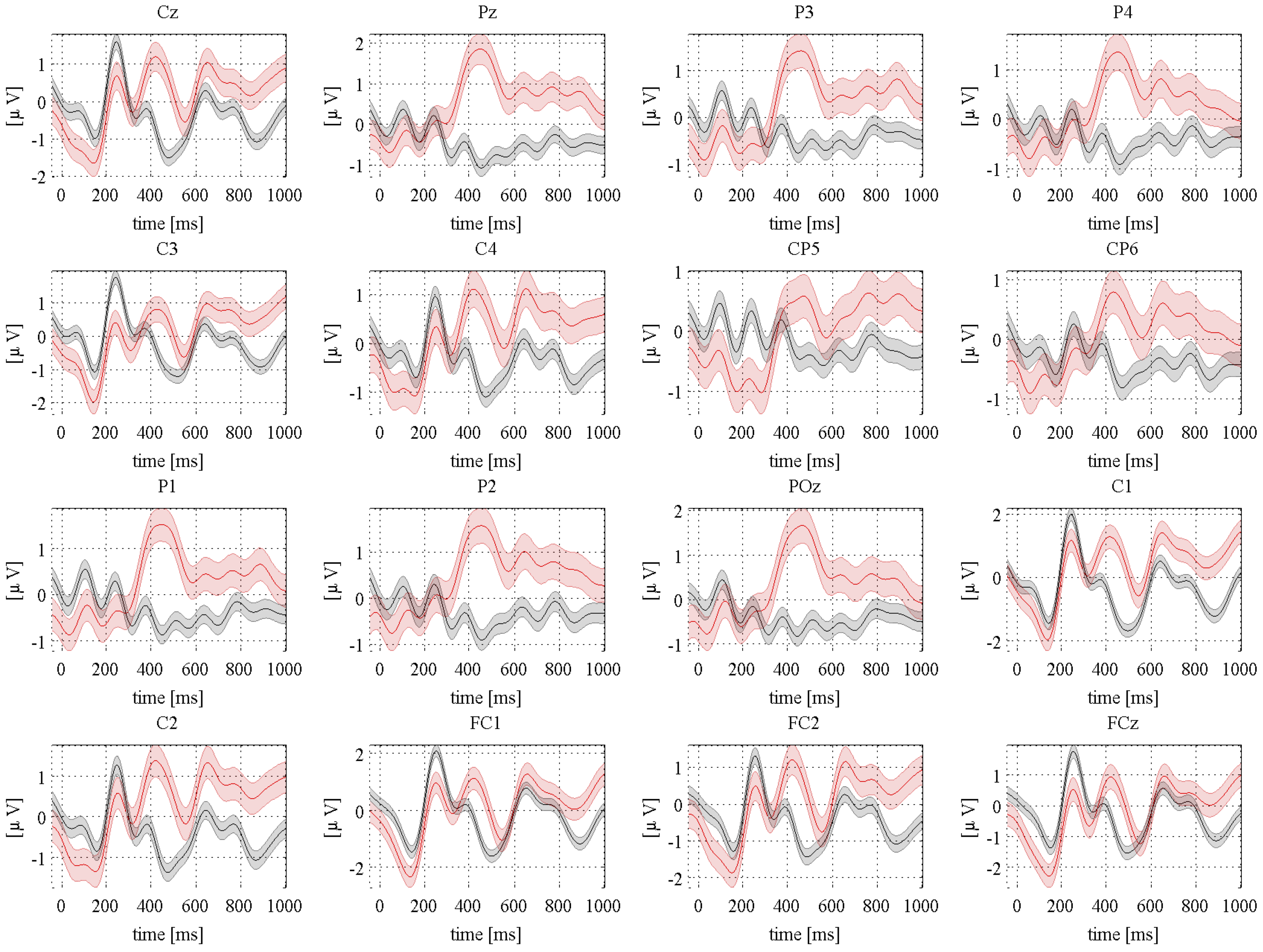}
	\caption{Grand mean averaged ERP of all participating subjects together. Each panel depicts responses from each electrode used in the study (see Table~\ref{tab:EEGconditions} for details). The red lines depict \emph{targets} and black \emph{non--targets}. The clear P300 responses could be seen in the range of $300 - 600$~ms. The further attentional modulation of \emph{target} responses extends till $1000$~ms.}
	\label{fig:EEGerp}
\end{figure}

The results of the conducted online tfBCI paradigm EEG experiment with the seven subjects are presented in Figure~\ref{fig:alleegauc} in form of matrices depicting ERP latencies with P300 response together with areas under the curve (AUC) feature separability analyses. We also present averaged topographic plots of the evoked responses at he latencies of the highest and lowest ERP separability in \emph{target vs. non--target} scenario. The highest average difference was found at $434$~ms (as calculated by AUC), which perfectly represented the P300 response peak as could be seen also in Figure~\ref{fig:EEGerp}, where \emph{target} and \emph{non--target} response are visualized separately for each electrode. Figure~\ref{fig:EEGerp} presents also a very interesting \emph{post--P300} attentional modulation which in the majority of chosen for our experiments electrodes had extended positive ERP modulation beyond the classical P300 peak in a range beyond $300-600$~ms and up to $1000$~ms. 

The online tfBCI accuracies (as obtained with SWLDA classifier) of the all seven participating subjects are summarized in Table~\ref{tab:EEGconditions}.
All the seven subjects scored well above the chance level of $25$\%. Four out of the seven subjects reached $100$\% accuracy based on $10$ ERP responses averaging, which is a very good outcome of the proposed tfBCI prototype. Based on the obtained accuracies we calculated, to allow simply comparison of the proposed tfBCI paradigm with other published approaches, the ITR scores which were in the range from $1.04$ bit/min to $10.00$ bit/min (see Table~\ref{tab:ITR}). The ITR was calculated as follows,
\begin{equation}
	ITR = V \cdot R,
\end{equation}
where $V$ is the classification speed in selections/minute ($5$~selections/minute in this case) and $R$ stands for the number of bits/selection calculated as, \begin{equation}
R=\log_{2}N + P \cdot \log_{2}P + (1 - P) \cdot \log_{2}\left(\frac{1 - P}{N - 1}\right),
\end{equation}
with $N$ being a number of classes (four in this study); and $P$ the classification accuracy (see Table~\ref{tab:eegaccuracy}). The ITR scores of the subjects in our study are summarized in Table~\ref{tab:ITR} and they shall be considered as good outcomes in comparison of the state--of--the--art BCI~\cite{bciBOOKwolpaw}.

\begin{table}
	\caption{Ten trials averaging classification BCI accuracy (note, this is not binary P300 classification result but resulting spelling result with a theoretical chance level of $25\%$) in tactile--force task using the classical SWLDA classifier~\cite{krusienski2006}.}\label{tab:eegaccuracy}
	\begin{center}
	\begin{tabular}{|c|c|}
	\hline
	Subject number 	& Online BCI experiment SWLDA accuracy \\
	\hline \hline
	$\#1$			& $100\%$ \\
	$\#2$			& $~75\%$ \\
	$\#3$			& $100\%$ \\
	$\#4$			& $100\%$ \\
	$\#5$			& $~50\%$ \\
	$\#6$			& $100\%$ \\
	$\#7$			& $~50\%$ \\
	\hline
	{\bf Average:}	& $\bf 82.1\%$ \\
	\hline
	\end{tabular}
	\end{center}
\end{table}

\begin{table}
	\caption{Ten trials averaging classification accuracy based ITR results (see Table~\ref{tab:eegaccuracy}).}\label{tab:ITR}
  	\begin{center}
	\begin{tabular}{|c|c|}
	\hline
	Subject number 	& ITR scores \\
	\hline \hline
	$\#1$			& $10.00$~bit/min \\
	$\#2$			& $~3.96$~bit/min \\
	$\#3$			& $10.00$~bit/min \\
	$\#4$			& $10.00$~bit/min \\
	$\#5$			& $~1.04$~bit/min \\
	$\#6$			& $10.00$~bit/min \\
	$\#7$			& $1.04$~bit/min \\
	\hline
	{\bf Average:}	& $\bf 6.58$~bit/min \\
	\hline
	\end{tabular}
	\end{center}
\end{table}

\section{Conclusions}

This case study demonstrated results obtained with a novel four--commands and tactile--force tfBCI paradigm developed and validated in experiments with seven healthy subjects. 
The experiment results obtained in this study confirmed the validity of the tfBCI application.

The EEG experiment with the paradigm has confirmed that tactile--force stimuli can be used easily (without any prior training) and successfully $1.04$~bit/min to $10.00$~bit/min for online case using SWLDA classifier. 

The results presented offer a step forward in the development of novel neurotechnology application. The current paradigm obviously needs still improvements and modifications to implement online with faster ISI and lower averaging rate necessary to improve the EEG features separability. These needs determine the major lines of study for future research. However, even in its current form, the proposed tfBCI can be regarded as a practical solution for LIS patients (locked into their own bodies despite often intact cognitive functioning), who cannot use vision or auditory based interfaces due to sensory or other disabilities.

We plan to continue this line of the tactile--force  BCI research in order to further optimize the feature extraction, signal processing and machine learning (classification) methods. Next we will test the paradigm with the LIS patients in need for BCI technology. 

\section*{Author Contributions}

Programmed the tactile--force stimulus generating and delivering interface: SK, DA, TMR. Performed the EEG experiments: SK, TMR. Analyzed the data: SK, TMR. Conceived the concept of the tactile--force BCI: TMR. Supported the project: SM. Wrote the paper: SK, TMR.

\section*{Acknowledgment}

This research was supported in part by the Strategic Information and Communications R\&D Promotion Program no. 121803027 of The Ministry of Internal Affairs and Communication in Japan.


\newpage

\newpage

\newpage

\newpage

\newpage
\begin{figure}
	\centering
	\vspace{-1cm}
	\includegraphics[width=0.65\linewidth]{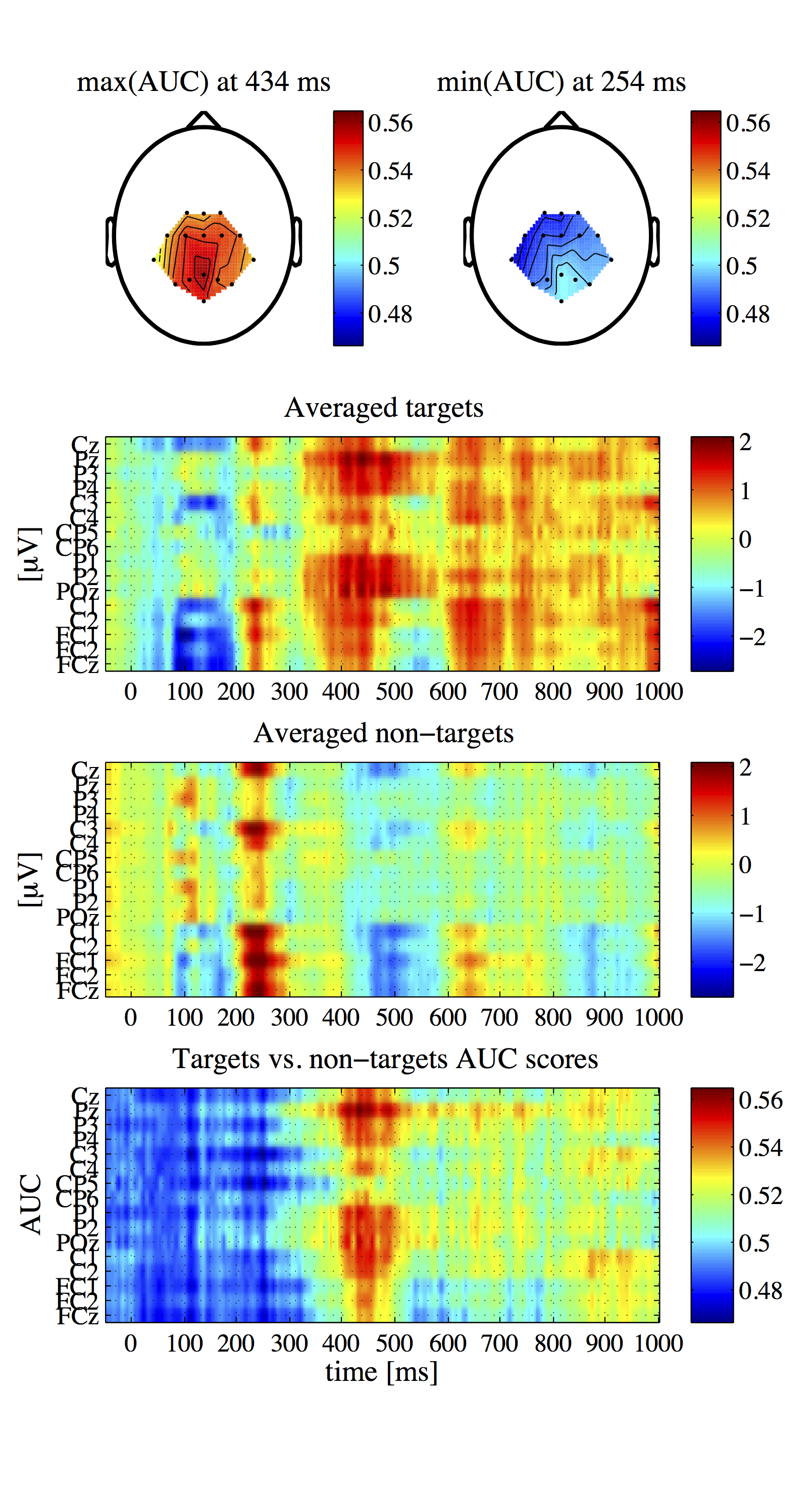}
	\vspace{-1cm}
	\caption{Grand mean ERP and AUC scores leading to final classification results of the participating seven subjects. The top panels represent head topographic plots of the {\em target} versus {\em non--target} area under the curve (AUC) scores (AUC is a measure commonly used in machine learning intra-class discriminative analysis and $\mbox{AUC} > 0.5$ confirms usually features separability). The top left panel represents a latency of the largest difference as obtained from the data displayed in the bottom panel of the figure. The top right panel depicts the smallest AUC latency. Those topographic plots also show the electrode positions. All the electrodes received similar AUC values (red) supports the initial electrode placement in the conducted tfBCI EEG experiments. The second panel from top represents averaged EEG responses to the {\em target} stimuli (P300 response in the range of $400$-$800$~ms). The third panel from top represents averaged EEG responses to the {\em non--target} stimuli (no P300 response). Finally, the bottom panel depicts the AUC of {\em target} versus {\em non--target} responses (P300 response latencies could be again easily identified here by red color--coded values).}
	\label{fig:alleegauc}
\end{figure}

\newpage
\begin{figure}
	\centering
	\vspace{-1cm}
	\includegraphics[width=0.65\linewidth]{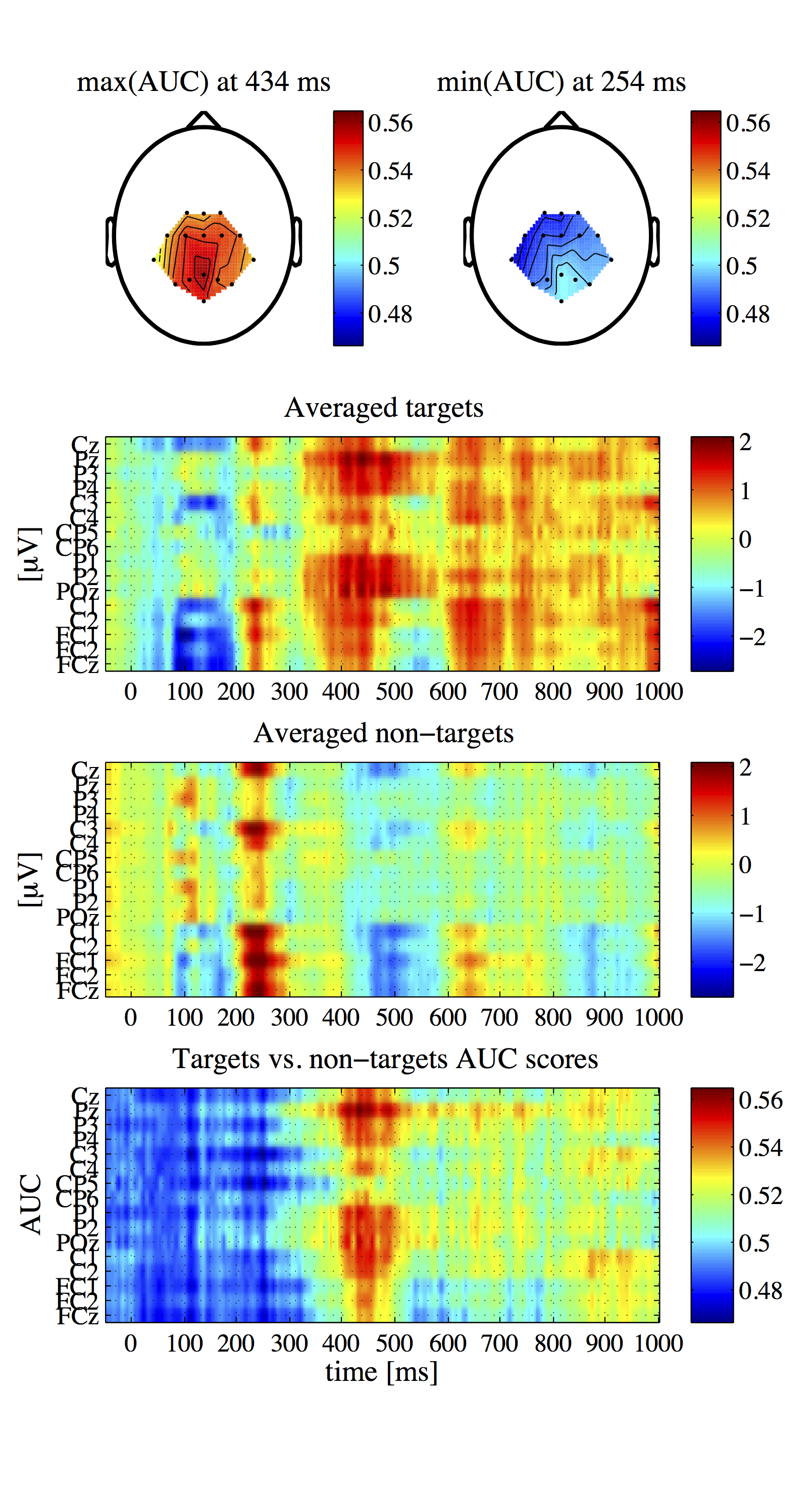}
	\vspace{-1cm}
	\caption{Grand mean ERP and AUC scores leading to final classification results of the participating seven subjects. The top panels represent head topographic plots of the {\em target} versus {\em non--target} area under the curve (AUC) scores (AUC is a measure commonly used in machine learning intra-class discriminative analysis and $\mbox{AUC} > 0.5$ confirms usually features separability). The top left panel represents a latency of the largest difference as obtained from the data displayed in the bottom panel of the figure. The top right panel depicts the smallest AUC latency. Those topographic plots also show the electrode positions. All the electrodes received similar AUC values (red) supports the initial electrode placement in the conducted tfBCI EEG experiments. The second panel from top represents averaged EEG responses to the {\em target} stimuli (P300 response in the range of $400$-$800$~ms). The third panel from top represents averaged EEG responses to the {\em non--target} stimuli (no P300 response). Finally, the bottom panel depicts the AUC of {\em target} versus {\em non--target} responses (P300 response latencies could be again easily identified here by red color--coded values).}
	\label{fig:alleegauc}
\end{figure}

\newpage

\end{document}